\documentclass[a4paper]{article}
\usepackage[margin=1in]{geometry}
\usepackage{cite}
\usepackage{graphicx}
\usepackage{verbatim}
\usepackage{amsmath}
\usepackage{array}
\usepackage{subcaption}
\usepackage{url}
\usepackage{color}

\begin{document}

\title{Formal Verification of Control Systems' Properties with Theorem Proving}

\author{Dejanira Araiza-Illan and Kerstin Eder\\Dept. Computer Science\\University of Bristol\\
Bristol, UK \\\{Dejanira.Araizaillan, Kerstin.Eder\}@bristol.ac.uk\\\\
Arthur Richards\\Dept. Aerospace Engineering\\University of Bristol\\
Bristol, UK\\Arthur.Richards@bristol.ac.uk}
\date{}

\maketitle

\begin{abstract}
This paper presents the deductive formal verification of high-level properties of control systems with theorem proving, using the Why3 tool. Properties that can be verified with this approach include stability, feedback gain, and robustness, among others. For the systems, modelled in Simulink, we propose three main steps to achieve the verification: specifying the properties of interest over the signals within the model using Simulink blocks, an automatic translation of the model into Why3, and the automatic verification of the properties with theorem provers in Why3. We present a methodology to specify the properties in the model and a library of relevant assertion blocks (logic expressions), currently in development. The functionality of the blocks in the Simulink models are automatically translated to Why3 as `theories' and verification goals by our tool implemented in MATLAB. A library of theories in Why3 corresponding to each supported block has been developed to facilitate the process of translation. The goals are automatically verified in Why3 with relevant theorem provers. A simple first-order discrete system is used to exemplify the specification of the Simulink model, the translation process from Simulink to the Why3 formal logic language, and the verification of Lyapunov stability.
\end{abstract}

\section{Introduction}
The use of formal methods allows the production of reliable and certified trustworthy autonomous systems~\cite{Simmons,Fisher} in a more automatic manner, facilitating the processes of verification and validation, and introducing methodologies for systems design towards verification. Verification has become essential for safety-critical autonomous systems. Autonomous systems have different compositional levels~\cite{Fisher}: agent (high-level planning and decision making), control (neural networks, controllers or control systems) and hardware implementation. Many proposed modelling formalisms and formal verification examples deal with the first layer only~\cite{Dixon,Webster,Siri}, or the first and second combined in a high-level abstraction (e.g., hybrid systems). For the latter, some approaches use theorem proving~\cite{Keymaeraa}. Other approaches translate the systems to decidable automata models and apply model checking~\cite{mc2002,MC2011,faults}, but these are not easily applicable to real-valued operations due to problems with scalability. Additionally, optimisation theory has been used to verify functionality and robustness of controllers for autonomous systems~\cite{Wang,Henrion}. 

The design of control systems typically begins with formal analysis followed by numerical implementation in a simulation tool like Simulink.  Numerical simulations then test for correct behaviour before the implementation is deployed.  In some cases, automatic code generation~\cite{Geneauto} is used to derive code directly from the simulation model.  In this paper, we propose the application of deductive formal verification methods on the implementation of the controller in Simulink.  This allows greater confidence in the correctness of the model with respect to its requirements.  In particular, the paper provides a way to translate Simulink models into the Why3 logic language to automatically prove results such as decrease of a Lyapunov function. One of the future aims of proposed approach is to verify the correctness of controllers based on an optimiser; for example, predictive controllers for UAV guidance~\cite{Arthur1,Arthur2}.

Why3 is a free and open source tool that interfaces with different theorem provers, particularly their Satisfiability Modulo Theory (SMT) solvers.  SMT solvers are extensions of SATisfiability (SAT) solvers that can accommodate real numbers, integers, and other domains (theories), and are thus well-suited to prove control systems properties~\cite{Metitarski}. Automatic theorem provers currently supported by Why3 include Alt-Ergo, CVC3, CVC4, E-prover, Gappa, Simplify, SPASS, Vampire, veriT, Yices and Z3, along with the interactive provers Coq and PVS~\cite{Why3b}. An SMT solver automatically computes the satisfiability of a logic formula based on a range of provided axioms and definitions.

The closest prior work to this paper in implementation is Simcheck~\cite{Simcheck}, which uses SMT solvers to perform type checking on a Simulink model with custom annotations. Stability based on Nichols plots has been verified for a system modelled in Simulink using the MetiTarski theorem prover~\cite{Metitarski}. Other approaches generate annotated C code from the model, then apply theorem proving to verify stability properties~\cite{Jobredeaux,Herencia}. Functional correctness of auto-generated code from Simulink models has also been formally verified through theorem provers~\cite{Clawz,Circus,Sparkfromsimulink,implementations} and model checking~\cite{Staats}.   

The main novel contributions in this paper are:
\begin{itemize}
\item A methodology to specify high-level properties for control systems in Simulink, supplemented by our custom blocks.
\item An automatic translation tool from Simulink to Why3, for formal verification of the properties. 
\end{itemize}
Our aim in pursuing this approach is to ease the inclusion of verification at design time, by incorporating it at the block diagram level where system interconnections and insight are best expressed.

In our proposed approach, higher-order logic requirements are incorporated and verified as Simulink signals~\cite{Hardware}. Note that this enables the same requirements to be tested by numerical simulation. This approach is partly inspired by the Open Verification Library (OVL)~\cite{OVL}, a library of modules that act as property checkers and can be placed into a hardware design where they are connected to regular modules using the circuit signals. For example, our new custom `Require' block, containing the built-in `Assert' block, incorporates pre- and post-conditions into the Simulink model. Then the automatic translation process to Why3 identifies the `Require' blocks as goals for theorem proving.  Meanwhile, each other block in the model is translated into the axioms it asserts relating its input and output signals. Finally, Why3 is invoked to run a chosen SMT solver on the translated `theory' and prove the goals, or otherwise.   

The rest of the paper presents relevant aspects of the Why3 language, and our proposed approach. Section~\ref{assertions} introduces our developed assertion blocks, particularly the `Require' block. Section~\ref{Why3syntax} presents a review of the Why3 logic language components. Section~\ref{translation} describes the translation process from Simulink to Why3, based on our developed theories in Why3 for different blocks. The translation is exemplified by a discrete system. Section~\ref{verification} explains the verification of Lyapunov stability for the same discrete system. The presented approach is discussed in the same section. Finally, the conclusion is presented in Section~\ref{conclusion}. 

\section{Assertion Blocks}\label{assertions}

We are developing a set of assertion blocks analogous to OVL for hardware verification~\cite{OVL}, to add to Simulink models to test and verify high-level requirements through specifications (as assertions) over the signals in the models. The OVL was developed to facilitate the addition of verification conditions (predefined logic expressions that can act as assertions, e.g.\ `event X always happens', assumptions or coverage points) to any hardware design, for assertion-based verification~\cite{assertions} or formal verification. The OVL checkers, written in Hardware Description Languages, receive names of the signals of interest as inputs, and are being monitored at simulation time when assertion violations are being recorded. The same assertions can serve as goals for formal verification.

The new blocks represent a range of assertions (logic expressions) to describe properties of interest related to control systems over signals, and others refer to structure for the assertions. Available Simulink assertion and logic blocks like `Compare To Zero', `Compare To Constant' and `Assertion' are also used as part or in conjunction with the new assertion blocks. The `Require' block denotes a requirement in a Hoare triple form~\cite{Hoare}
 \begin{equation}
\{\text{preconditions}\} \rightarrow [ \text{model} ] \{\text{postconditions}\},
\end{equation}
where the logic expressions for the preconditions and postconditions are translated from the assertion blocks connected to the `Require' block. The translator automatically produces the verification goals for the theorem prover from these blocks. The `Require' block is an enabled subsystem with an `Assert' block inside (Fig.~\ref{requireinside}). They use native Simulink and MATLAB formats, thus allowing numerical testing in simulation and coverage assessment. 

\begin{figure}[!t]
\begin{center}
\begin{subfigure}{0.33\textwidth}
                \includegraphics[width=1.5in]{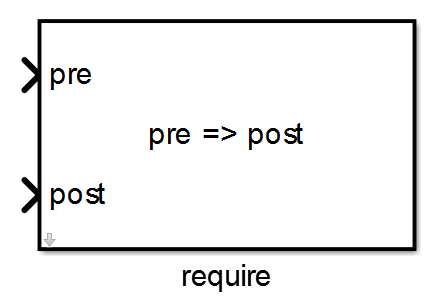}
                \caption{The `Require' Block.}
                \label{subsystem}
        \end{subfigure}%
\begin{subfigure}{0.33\textwidth}
                \includegraphics[width=1.5in]{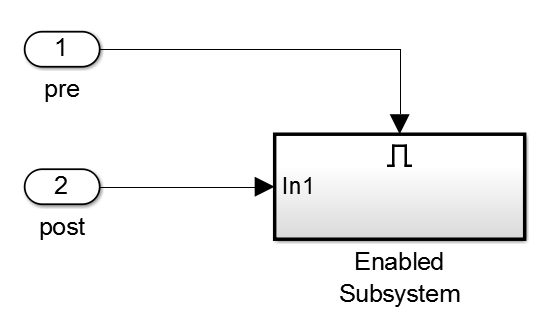}
                \caption{`Enabled Subsystem' Inner Blocks.}
                \label{subsystem}
        \end{subfigure}%
\begin{subfigure}{0.33\textwidth}
                \includegraphics[width=1.5in]{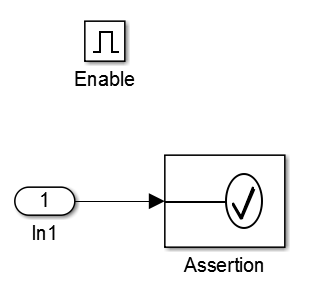}
                \caption{The `Require' Block and Inner Components.}
                \label{subsystem}
        \end{subfigure}%
        \end{center}
\end{figure}

\begin{figure}[!t]
\centering
\includegraphics[width=3.5in]{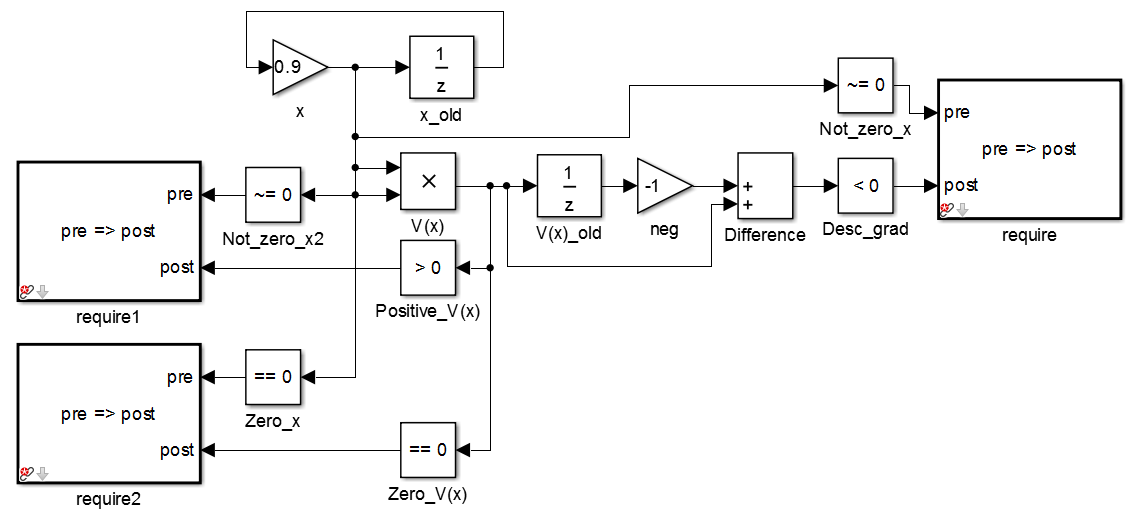}
\caption{A First-Order System Specified with Assertion Blocks in Simulink.}
\label{systemcomplex}
\end{figure}

In the first-order system of Fig.~\ref{systemcomplex}, we have implemented a Lyapunov function $V(x)=x^2$ for the analysis of stability. Three `Require'  blocks (`require', `require1' and `require2') describe stability properties parametrised by signals in the model: 
\begin{itemize}
\item `require': $x\neq 0 \rightarrow V(x(k))-V(x(k-1))<0$. If the signal $x$ is different from 0, the signal leaving the `Difference' block should be negative.
\item `require1': $x\neq 0 \rightarrow V(x)>0$. If the signal $x$ is different from 0, then the signal $V(x)$ should be positive. 
\item `require2': $x=0 \rightarrow V(x)=0$. If the signal $x$ is 0, then the signal $V(x)$ should be zero.   
\end{itemize}

More complex assertions are built-up from multiple basic blocks, which guarantees their translatability into Why3 (using the predefined theories for basic blocks) and preserves the modularity in Simulink. Properties can be verified in independent sections of blocks, or even within blocks inside subsystems. Individual proofs  can be assembled into a hierarchical proof structure, to provide conclusions at a higher level from the sequence and relations of the lower-level proofs.

\section{Review of Why3 Syntax} \label{Why3syntax}

The Why3 logic language is based on polymorphic types and first-order logic, although higher-order logic syntax is allowed~\cite{Why3b,Why3man}. The WhyML language targets the formal verification of programs, by including loops and other control structures. The Why3 logic language is sufficient to express and verify block diagrams from Simulink due to their declarative nature. The translation from Simulink to Why3 logic expressions is based on the following Why3 components: 
\begin{itemize}
\item Theories: blocks of logic expressions with types, functions, predicates, axioms, lemmas and verification goals that can be used or cloned in a modular manner. More complex theories can be formed from basic theories, inheriting the axioms, lemmas, functions and declarations. They contain the components for a proof. 

\begin{verbatim}
theory <Name>
...
end
\end{verbatim}

\item Functions: operations over data, that can be recursive and accept different types (e.g., integers, real, Boolean). The parameters are expressed in curried syntax, where a function $f(x)$ becomes \verb+f x+. 

\begin{verbatim}
function <name> <inputs> <types> : <outputs> <types>
\end{verbatim}

\item Axioms and lemmas: logic expressions to help a proof. 

\begin{verbatim}
axiom/lemma <name>: <logic expression>
\end{verbatim}

\item Goals: logic expressions to prove. 

\begin{verbatim}
goal <Name>: <logic expression>
\end{verbatim}

\item Theories can be used (same symbols) and cloned (no reuse of symbols) in other theories. Some basic theories are available, including real numbers, integers, floating-point and lists. 

\begin{verbatim}
use import <file_name>.<Name>

clone <file>.<Name> as <New_name> with function 
<name> = <new_name>
\end{verbatim}
\end{itemize}

We have developed a library of theories that provide axioms applicable to the input-output behaviour corresponding to popular Simulink blocks, including arithmetic and logic, to facilitate the translation into Why3. The theories are cloned following the components of a block diagram, and the axioms within each theory are parametrised according to signal names and specific values (constants, input ranges or initial conditions). The theory developed for the `Product' block is shown in Fig.~\ref{theory}. In the future, additional linear algebra and control systems related theories will be developed as needed in order to verify more complex systems and properties, as in~\cite{Herencia}. 

\begin{figure}[!t]
\begin{center}
\begin{verbatim}
theory Product_int
 use import int.Int  
 use import real.RealInfix
 
 function in1 int: real
 function in2 int: real
 function out1 int: real
 
 axiom v: forall k:int. out1 k = in1 k *. in2 k
 axiom c1: forall k:int. in1 k >. 0.0 /\ in2 k >. 0.0 -> out1 k >. 0.0
 axiom c2: forall k:int. in1 k <. 0.0 /\ in2 k <. 0.0 -> out1 k >. 0.0
end
\end{verbatim}
\end{center}
\caption{Theory for the `Product' Block.}
\label{theory}
\end{figure}

In the \verb+Product_block + theory, the theories of the integer and real numbers are imported first. Then, two functions are used for the inputs and another for the output. They receive an integer (time sample) and produce a real (the signal value at that time). Three axioms describe the multiplication and some sign properties: for all sampled times, \verb+forall k:int.+, the output signal is equal to the multiplication of the inputs (\verb+axiom v+);  the multiplication of two positive numbers produces a positive number (\verb+axiom c1+); and the multiplication of two negative numbers produces a positive number (\verb+axiom c2+). The algebraic and logic symbols are used with a dot when a theory mixes integer and real numbers (i.e., both theories are imported in the same theory), whereas if a theory only uses integer or real numbers, the symbols are used without the dot~\cite{Why3man}. 

The theory developed for the `Compare To Zero' block with condition case $\sim = 0$ (not equal to zero) is shown in Fig.~\ref{theory2}.  

\begin{figure}[!t]
\centering
\begin{verbatim}
theory CompareToZero_neq_int
 use import int.Int  
 use import real.RealInfix
 use import bool.Bool
 
 function in1 int: real
 function out1 int: bool
 
 axiom v1: forall k:int. out1 k = True -> in1 k <>. 0.0
 axiom v2: forall k:int. out1 k = False -> in1 k = 0.0 
end
\end{verbatim}
\caption{Theory for the `Compare To Zero' Block, for the $\sim = 0$ Case.}
\label{theory2}
\end{figure}

In the \verb+CompareToZero_block_ne0 + theory, the theories of the integer, real and Boolean numbers are imported first. Then, a function is defined for the input and another for the output. The input receives an integer (time sample) and produces a real value (the signal value at that time), whereas the output produces a Boolean value (\verb+True + or \verb+False+). Two axioms describe the conversion from a real valued signal (input) to a Boolean (output), according to the comparison criterion: if the input value is different to zero (\verb+axiom v1+), the output is \verb+True+; and if the input value is zero the output is \verb+False+. 

\section{Translation from Simulink to Why3} \label{translation}

The automatic translation process from Simulink to Why3 converts the signals and blocks into higher-order logic predicates~\cite{Hardware}. For example, the functionality of a `Product' block over two inputs and an output signal in discrete intervals of time ($k$) is expressed in higher-order logic~\cite{Hardware} as
\begin{equation}
Product(in_1,in_2,out_1) \,\equiv \, \forall k. \,out_1(k)=in_1(k)*in_2(k).
\end{equation}

The proposed and implemented procedure for the translation of systems as block diagrams in Simulink into Why3 language consists of the following steps, executed in automatically in MATLAB when calling the translator function:
\begin{enumerate}
\item  Identification of the specification blocks from the rest of the model. For example, `Require' blocks and their associated preconditions and postconditions, where the latter two are directly connected to the respective labelled inputs of the `Require' blocks. 

\item The Abstract Syntax Tree (AST)~\cite{PVS,Simcheck} of the model is computed automatically. We implemented a program that identifies the predecessors and successors of each block, using the block manipulation functions available in MATLAB. Each AST is translated into a theory in Why3, named after the model to be verified. Relevant numerical theories including real numbers, integers and Boolean expressions are imported. 

\item Then, every signal that leaves a block is identified, named and translated to real or Boolean functions. Functions that receive an integer represent discrete signals (i.e., signals of the form $y(k)$ where $k$ is the time step):

\begin{verbatim}
function <block_name>_op<n> int: <real/bool>
\end{verbatim}
where \verb+<n>+ is the corresponding output port (in case the block has more than one), \verb+int+ indicates if the signal is discrete, and \verb+real/bool+ indicates if the signal is numeric or Boolean, respectively. In this paper, all the signals in Simulink are scalar and discrete, although the translation and verification will be extended to continuous time in the future. 

\item Afterwards, the theories corresponding to all the blocks in the system are cloned automatically, with parametrisation (e.g., gains, constants) as added axioms. The cloning of some structural specification blocks like `Require' is excluded, as the main purpose of these blocks in the translation is to give structure to the verification goals (e.g., Hoare triple inference form).  

\item Finally, the verification goals are added following structures like the Hoare triple form from the `Require' blocks. The preconditions and postconditions are added as desired Boolean signal outcomes in the Hoare triple: 

\begin{verbatim}
goal <Name>: <time>. <precond.>  -> <postcond.>
\end{verbatim}
\end{enumerate}

If the Simulink model has nested generic subsystems (i.e., grouping blocks with the purpose of providing modularity), each subsystem is translated into a theory from the computation of its internal AST and verification goals, using the previously described procedure. These theories are then cloned into other theories, following the flow diagrams of upper levels. 

\begin{figure}[!t]
\centering
\includegraphics[width=3.5in]{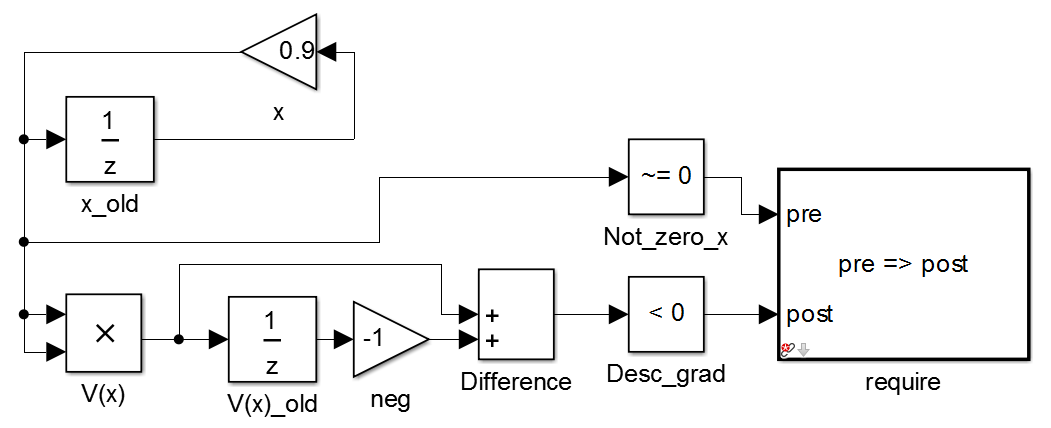}
\caption{Reduced First-Order System in Simulink.}
\label{system}
\end{figure}

Consider the first-order discrete system in Simulink shown in Fig.~\ref{system}, a simplification of Fig.~\ref{systemcomplex}. The translator produces the theory in Fig.~\ref{alltheory} automatically. 

\begin{figure*}[!t]
\centering
\scriptsize
\begin{verbatim}
theory M_firstorder
 use import int.Int
 use import real.RealInfix
 use import bool.Bool

 function difference_op1 int: real
 function vx_op1 int: real
 function vx_old_op1 int: real
 function neg_op1 int: real
 function x_op1 int: real
 function x_old_op1 int: real
 function not_zero_x_op int: bool
 function desc_grad_op int: bool

  clone simulink.Sum_int as Difference with function in1 = vx_op1, function in2 = neg_op1, function out1 = difference_op1

  clone simulink.Product_int as Vx with function in1 = x_op1, function in2 = x_op1, function out1 = vx_op1

  clone simulink.UnitDelay_int as Vx_old with function in1 = vx_op1, function out1 = vx_old_op1

  clone simulink.Gain_int as Neg with function in1 = vx_old_op1, function out1 = neg_op1
  axiom neg_gain: Neg.gain = -.1.000000

  clone simulink.Gain_int as X with function in1 = x_old_op1, function out1 = x_op1
  axiom x_gain: X.gain = 0.900000
 
  clone simulink.UnitDelay_int as X_old with function in1 = x_op1, function out1 = x_old_op1

  clone simulink.CompareToZero_neq_int as Not_zero_x with function in1 = x_op1, function out1 = not_zero_x_op

  clone simulink.CompareToZero_l_int as Desc_grad with function in1 = difference_op1, function out1 = desc_grad_op

  goal G1 : forall k: int. not_zero_x_op k = True -> desc_grad_op k = True
end
\end{verbatim}
\caption{Theory for the First Order System in Fig.~\ref{system}.}
\label{alltheory}
\end{figure*}

The name of the theory, in the first line, corresponds to the name of the model file in Simulink. Then, the theories of the integer, real and Boolean numbers are imported. Functions corresponding to the signals that come out of the blocks are added to the theory, specifying if they correspond to real or Boolean signals, and following the convention for the naming described before. 

Subsequently, the theories of the different blocks in the Simulink model are cloned and parametrised from our library (contained in a file named `simulink') mentioned in Section~\ref{Why3syntax}. The order of the blocks is given by the Simulink automatic identification numbering: the `Sum' block, the `Product' block (the Lyapunov function), the `Unit Delay' block of the Lyapunov function, the `Gain' block to achieve the subtraction of the values of the Lyapunov function combined with the `Sum' block, the `Gain' block and the `Unit Delay' of the system.  When a theory from the library is cloned, its name is changed to the name of the block. The input and output functions of the new theory are renamed according to the signals connected to the respective block. Examples of value parametrisation in the cloning are the axioms corresponding to the gains \verb+Neg_gain+ and \verb+x_gain+, for the `Gain' blocks in the model. 

The theories corresponding to the comparison blocks (the two `Compare To Zero' blocks) are cloned considering the parameters assigned in the model, i.e., calling the respective theory that corresponds to a comparison of any of the possible types $\{==,\sim =,>,<,>=,<=\}$. Finally, the verification goal is added by connecting the precondition and postcondition following the structure of the Hoare triple, \verb+not_zero_x_op1 k = True -> desc_grad_op1 k = True+.

\section{Verification of a First-Order System}\label{verification}

Consider the same first-order linear discrete system as in Fig.~\ref{system}, for a single dimension (scalar),
\begin{equation}
x(k+1)=ax(k),
\end{equation}
and a metric for stability given by a quadratic Lyapunov function, that is always positive regardless of its input,
\begin{equation}
V(x)=x^2.
\end{equation}

A system like the one in our example is considered stable if its trajectory in time converges to a region or point in the state space, starting from different initial conditions. A way to prove stability of the system in Fig.~\ref{system} is the existence of a Lyapunov function that preserves its expected properties of positivity, and that will decrease (if the system is stable), stay in the same value (if the system is marginally stable), or increase (if the system is unstable). Stability is proved in our examples for different values of the gain $a$ computing the discrete gradient descent of the Lyapunov function (the difference between the value of the function in the current time interval and in the previous interval). 

The conditions to prove stability through the signals has been expressed in two assertions added as blocks to the Simulink model: when the signal $x$ is not zero (precondition), if the system is stable the gradient of the Lyapunov function should be negative (postcondition), 
\begin{equation}
x(k)\neq0 \rightarrow V(x(k))-V(x(k-1))<0
\end{equation}
as shown in Fig.~\ref{system}. Thus, a failure of proof is expected for the verification of the goal \verb+goal G1 + when the gain is $>=1$ or $<=-1$, as the implication would lead to True~$\rightarrow$~False, which is in turn False. The validity of the proof is expected when the gain is $<1$, $>-1$ or $=0$, as the first two lead to True~$\rightarrow$~True, and the later to False~$\rightarrow$~False, all implications resulting in True. These conclusions are derived from the truth table for the logical implication operation. 

\subsection{Assertion Checks in Simulation}

The stability assertions specified in the Simulink model have been checked in simulation, for different gain values in the system loop. If the system does not comply with the specified property, the `Assert' block inside the `Require' block will fire up a flag when running the Simulink model. The results for different gains are shown in Table~\ref{check}, conforming to our predictions: failure when the gain is $>=1$ and $<=-1$ (thus the system is unstable), and success in the checks when the gain is $<1$, $>-1$ (thus the system is stable), highlighted in the table. When the gain is $=0$, the `Assert' block inside the `Require' is not active and the check does not take place, due to the disabling action of the precondition in the `Enabled Subsystem', as the value of the signal $x(k)$ is 0 and thus the precondition is false. 

\begin{table}[!t]
\renewcommand{\arraystretch}{1.3}
\caption{Checks with Different Gains}
\label{check}
\centering
\begin{tabular}{|c|c|}
\hline
GAIN&CHECK RESULT\\ \hline \hline
-1.1&Fail \\ \hline
-1.0&Fail \\ \hline
\textbf{-0.5}&\textbf{Pass} \\ \hline
0.0&Not checked\\ \hline
\textbf{0.8}&\textbf{Pass}\\ \hline
\textbf{0.9}&\textbf{Pass}\\ \hline
\textbf{0.9999}&\textbf{Pass}\\ \hline 
1.0&Fail \\ \hline
1.1&Fail \\ \hline
2.0&Fail \\ \hline
\end{tabular}
\vspace{-0.1cm}
\end{table}

\subsection{Proofs in Why3}
The goal has been verified for different gains using the CVC3 SMT solver, compatible for theories with real and integer numbers, linear arithmetic, equalities and inequalities. The results of the verification are shown in Table~\ref{setp}. The highlighted results indicate proved stability (not including marginal), from the validity of both the precondition and postcondition, and the `Unknown' results show that stability could not be proved for those gains. These results match the assertion checks presented in Table~\ref{check}. CVC3 automatically computed the validity of the verification goals in less than 150 seconds, our specified time limit in Why3. 

\begin{table}[!t]
\renewcommand{\arraystretch}{1.3}
\caption{Verification with Different Gains}
\label{setp}
\centering
\begin{tabular}{|c|c|c|c|}
\hline
GAIN&G1: stability&Prec.&Postc.\\ 
(\verb+X.gain+\normalsize )&$x(k)\neq0 \rightarrow$&&\\
&$V(x(k))-V(x(k-1))<0$&& \\ \hline \hline
-1.1&Unknown&True&False \\ \hline
-1.0&Unknown&True&False \\ \hline
\textbf{-0.5}&\textbf{Valid}&\textbf{True}&\textbf{True} \\ \hline
0.0&Valid&False&False\\ \hline
\textbf{0.8}&\textbf{Valid}&\textbf{True}&\textbf{True}\\ \hline
\textbf{0.9}&\textbf{Valid}&\textbf{True}&\textbf{True}\\ \hline
\textbf{0.9999}&\textbf{Valid}&\textbf{True}&\textbf{True}\\ \hline 
1.0&Unknown&True&False \\ \hline
1.1&Unknown&True&False \\ \hline
2.0&Unknown&True&False \\ \hline
\end{tabular}
\end{table}

The case study presented in this paper showcases our approach to formally prove properties of interest, from the translation of Simulink into Why3. The translation is performed automatically for a subset of Simulink blocks. This subset is currently in expansion. Different theorem provers can be selected in Why3 for the proof, according to their capabilities. The main disadvantage of Why3 is the lack of feedback about the failure in the proof, compared to the production of counterexamples in model checking tools. Nevertheless, the tool chain can be extended to incorporate equivalent mechanisms for feedback.

\section{Conclusion} \label{conclusion}
This paper presented an approach to formally verify high level properties of interest of control systems (e.g., stability) as Simulink models in a more automatic manner in Why3 (using theorem proving tools). The Simulink model is specified or annotated with assertions in blocks from our proposed reference library. The Simulink models are translated automatically into the Why3 logic syntax, from the signals in time and the block connections (computing an abstract syntax tree). The translation process is helped by our library of theories in Why3 that correspond to the functionality of a set of Simulink blocks. The assertion blocks are translated into verification goals, and then proved by calling different theorem provers in Why3 according to their capabilities (theories). 

An example of the translation and verification was provided in the form of a first-order discrete system, coupled with a Lyapunov function as a metric for stability. The verification goals were proved or otherwise, according to the expected stability results for different parameters (gain and initial values), using the CVC3 theorem prover (or its embedded SMT solver). Assertion checks were performed over the Simulink model in simulation, to show the useful duality of our proposed specification approach. 

Immediate future work includes extending our approach to multidimensional problems by incorporating linear algebra theories in Why3. Other future work includes the expansion of our libraries of assertion blocks and theories for supported Simulink blocks, for other high-level properties of interest (e.g., robustness) and stability methods.  We are intending to apply our methodology and tools to a diverse range of systems.

\section*{Acknowledgment}

The work presented in this paper was supported by the EPSRC grant EP/J01205X/1 RIVERAS: Robust Integrated Verification of Autonomous Systems.

\bibliographystyle{plain}

\bibliography{references_paper}

\end{document}